\documentclass[lettersize,journal]{IEEEtran}
\usepackage{amsmath,amsfonts}
\usepackage{algorithmic}
\usepackage{algorithm}
\usepackage{array}
\usepackage[caption=false,font=normalsize,labelfont=sf,textfont=sf]{subfig}
\usepackage{textcomp}
\usepackage{stfloats}
\usepackage{url}
\usepackage{verbatim}
\usepackage{graphicx}
\usepackage{cite}
\usepackage{xcolor}
\begin{document}
\title{6G comprehensive intelligence: network operations and optimization based on Large Language Models}

\author{Sifan Long, Fengxiao Tang, Yangfan Li, Tiao Tan, Zhengjie Jin, Ming Zhao, and Nei Kato
\thanks{Sifan Long, Fengxiao Tang (corresponding author), Yangfan Li, Tiao Tan, Zhengjie Jin and Ming Zhao are with Central South University, China; Nei Kato is with Tohoku University, Japan.}}

\markboth{Journal of \LaTeX\ Class Files,~Vol.~14, No.~8, August~2021}%
{Shell \MakeLowercase{\textit{et al.}}: A Sample Article Using IEEEtran.cls for IEEE Journals}

\maketitle
\begin{abstract}
The sixth generation mobile communication standard (6G) can promote the development of Industrial Internet and Internet of Things (IoT). To achieve comprehensive intelligent development of the network and provide customers with higher quality personalized services. This paper proposes a network performance optimization and intelligent operation network architecture based on Large Language Model (LLM), aiming to build a comprehensive intelligent 6G network system. The Large Language Model, with more parameters and stronger learning ability, can more accurately capture patterns and features in data, which can achieve more accurate content output and high intelligence and provide strong support for related research such as network data security, privacy protection, and health assessment. This paper also presents the design framework of a network health assessment system based on LLM and focuses on its potential application scenarios, through the case of network health management system, it is fully demonstrated that the 6G intelligent network system based on LLM has important practical significance for the comprehensive realization of intelligence. 
\end{abstract}


\section{Introduction}
The intelligence of 6G networks can enhance the network's perception and cognitive abilities, ensure network security, and drive application innovation [1]. It has developed for many years since the maturity of Deep Learning technology. However, due to the limited capacity of small models in handling specific tasks, and even their difficulty in dealing with complex scenarios, these limitations have led to lower network intelligence. On the one hand, with the further development of Artificial Intelligence (AI) technology, Large Language Models (LLMs) based on Deep Learning have proven to possess powerful generalization and reasoning capabilities. These models are able to learn and extract language patterns and rules from vast amounts of textual data, enabling them to deeply understand and generate natural language. Their robust ``emergent" abilities guarantee their performance in providing powerful general task-solving and complex task-reasoning capabilities within complex systems. As shown in Figure 1, networks based on Large Language Models have achieved numerous applications, allowing for convenient handling of traditional issues faced in network operations and performance optimization, such as network monitoring, network health assessment, network analysis, and resource scheduling.
\begin{figure}[!t]
	\centering
	\includegraphics[width=3.5in]{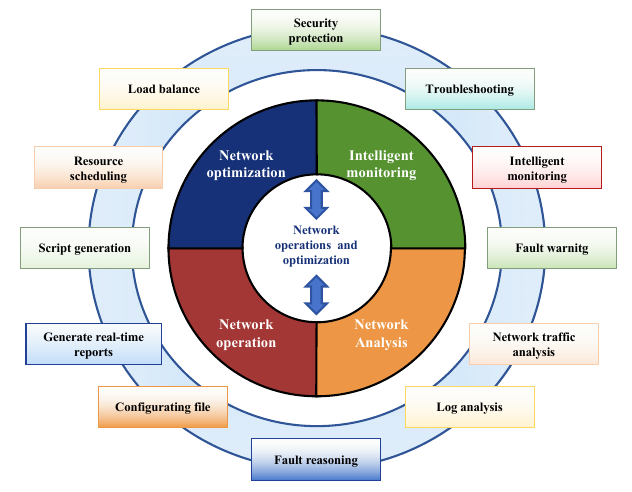}
	\caption{Application of Large Language Model in intelligent network operations and performance optimization.}
	\label{fig_a1}
\end{figure}

On the other hand, thanks to the advantages of the Transformer model [2] in parallelized computation, capturing long-distance dependencies, adapting to various tasks, and superior performance, research on Large Language Models [3] based on Transformer has rapidly developed in various application research fields, and it achieved remarkable progress in the field of Natural Language Processing. As shown in Figure 2, various multimodal tasks based on processing text, images, and videos have emerged one after another in the past ten years, and it has sparked great interest in the Artificial Intelligence community and the entire world. Currently, the main participants in the Artificial Intelligence industry are competing to develop their own proprietary LLM frameworks, so that they can be applied to their respective fields. Representative examples include OpenAI's GPT-3, Google's PALM, and Meta's LLaMA, whose parameters have exceeded billions, enabling them to adapt to fitting modeling in complex application scenarios. At the same time, these LLM frameworks are trained on a wide range of data sets from the Internet.  When the size of the model increases significantly, they show a strong generalization ability, enabling it to easily handle multimodal tasks. In recent years, driven by the research on 6G networks, there have been greater challenges in designing and optimizing communication protocols and general methods for 6G networks [4], such as low latency, high frequency band, high bandwidth, high transmission rate, and intelligence. 
Meanwhile, the network application architecture based on Large Language Models encompasses multiple crucial stages, including data preprocessing, model training, building and deployment, network application integration, as well as model optimization and iteration. These stages collectively form the complete lifecycle of Large Language Models in network applications, and drive the development of network intelligence and application innovation.
The LLM framework based on Transformer can effectively leverage its unique advantages in complex 6G network research, making 6G networks more intelligent. The advantages of Large Language Model in network operations and optimization are mainly reflected in its strong expression ability and generalization ability, as well as the ability to handle complex tasks. This is also the motivation of this paper.

\begin{figure}[!t]
	\centering
	\includegraphics[width=3.6in]{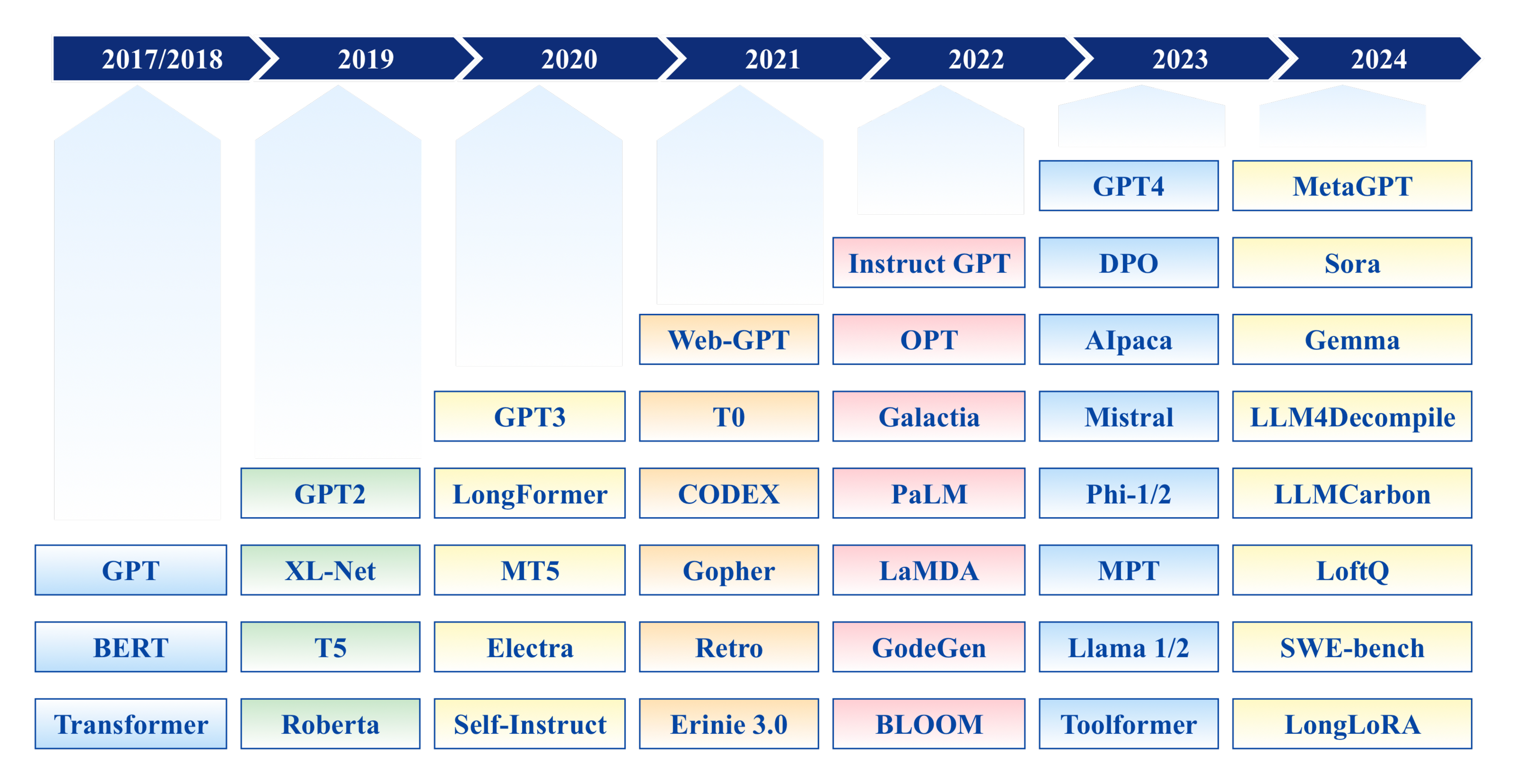}
	\caption{The development timeline of the typical representative LLM framework.}
	\label{fig_b1}
\end{figure}
\section{The background of Large Language Models and related applications in the network}
\subsection{Large Language Models}
Large Language Models (LLMs) have demonstrated extraordinary capabilities in Natural Language Processing (NLP) [5], including tasks such as translation, question answering, and text generation. From the initial representative GPT model to the latest Sora model as shown in Figure 2, the capabilities of LLM have been greatly improved. For example, Sora [6] can even generate video images that are difficult to distinguish between real and fake, pushing the research on applying LLM to video generation to another climax.  

Currently, there are three main types of LLM architectures that are widely used: encoder-only, decoder-only, and encoder-decoder. The first category is the ``encoder-only'' language model, which is good at text understanding because it allows information to flow in both directions in the text. The second category is the ``decoder-only'' language model, which is good at text generation because information can only flow from the left to the right of the text and effectively generate new words in an autoregressive manner. Finally, the ``encoder-decoder'' group combines the first two models to complete tasks that require understanding input and generating output, such as real-time translation. However, they are all built using the Transformer architecture as a basic component, the core of Transformer is the (self-)attention mechanism, which can capture long-term context information more effectively than recurrence and convolution mechanisms using GPUs. Therefore, the advent of the Transformer architecture marks the beginning of the era of modern Large Language Models.
The LLM has achieved many successful applications in the fields of fault diagnosis, image processing, protein structure prediction, drug research and development, autonomous driving technology, financial analysis, and climate research.

\subsection{Related Work of LLM in the Network}
Artificial intelligence agents based on multimodal Large Language Models are expected to revolutionize the research on traditional network applications, especially the deployment of LLM agents in 6G networks, which holds greater significance for the design of network structures and performance optimization. Therefore, intelligence based on multimodal LLMs has become a more prominent feature of 6G networks, and as a result, intelligent agents based on LLMs have achieved many important applications in various fields of network application research.
For instance, Lai \textit{et al.} [7] proposed a general framework for resource-efficient generative incentive mechanism in resource-constrained scenarios. They fully demonstrated their feasible solutions for handling issues related to network overhead, resource allocation, and task offloading in generative mobile edge networks through application cases. Furthermore, in the design of collaborative Cloud-Edge-Device network communication structures, the collaborative Cloud-Edge-Device methodology exhibits unique capabilities in generating personalized services. Chen \textit{et al.} [8] introduced NetGPT, which can handle cloud collaboration, providing end-users with personalized and inclusive network intelligence, while granting them privileged operations for real-time access to generative services. Additionally, in terms of 6G network performance optimization, Xu \textit{et al.} [9] proposed a split learning system for LLM in 6G networks to reduce interaction latency and better protect user privacy. They leveraged the collaboration between mobile devices and edge servers, distributing multiple LLMs with different roles across mobile devices and edge servers to efficiently collaborate in executing user-agent interaction tasks. This system can offload complex tasks to global LLMs running on edge servers, effectively improving the communication efficiency of the entire network.
As we discussed earlier, the Large Language Models have demonstrated remarkable capabilities in areas such as Natural Language Processing and Computer Vision (CV), revolutionizing the development concept of artificial intelligence and potentially reshaping our future. However, given the multimodal nature and structural complexity of the cloud, cloud-based deployments may face critical challenges, including long response times, high bandwidth costs, and violations of data privacy. Nevertheless, 6G Mobile Edge Computing (MEC) systems can effectively address these pressing issues [10].
Taking into account the inherent resource constraints at the edge, Lin \textit{et al.} [11] have discussed various cutting-edge techniques, including split learning/inference, parameter-efficient fine-tuning, quantization, and parameter-sharing inference, to facilitate the efficient deployment of LLMs. It is believed that with the further development of AI, the LLM-based network applications will become more widespread, further driving human society towards comprehensive intelligence.

Most of the above summarized works merely focus on solving the fundamental problems faced by network optimization in specific scenarios, encompassing antenna technology, radio frequency technology, base station technology, and more, to enable efficient communication networks with low-latency and low-bandwidth requirements. Furthermore, the research and development of high-frequency technologies such as millimeter-wave and terahertz are crucial for 6G networks to achieve high bandwidth and low latency. However, these technologies are still in their infancy, with numerous technical obstacles to overcome. In the context of complex network operation and performance optimization applications, designing a comprehensive and unified universal framework as a solution for efficient operation and performance optimization of 6G networks remains a formidable challenge. Therefore, to address this challenge, this paper proposes an intelligent network operation and performance optimization architecture that can tackle some common technical issues encountered in network operation and performance optimization. This proposal holds significant practical implications for realizing efficient communication in 6G networks.

\section{Intelligent network architecture}
\begin{figure*}[!t]
	\centering
	\includegraphics[width=7in]{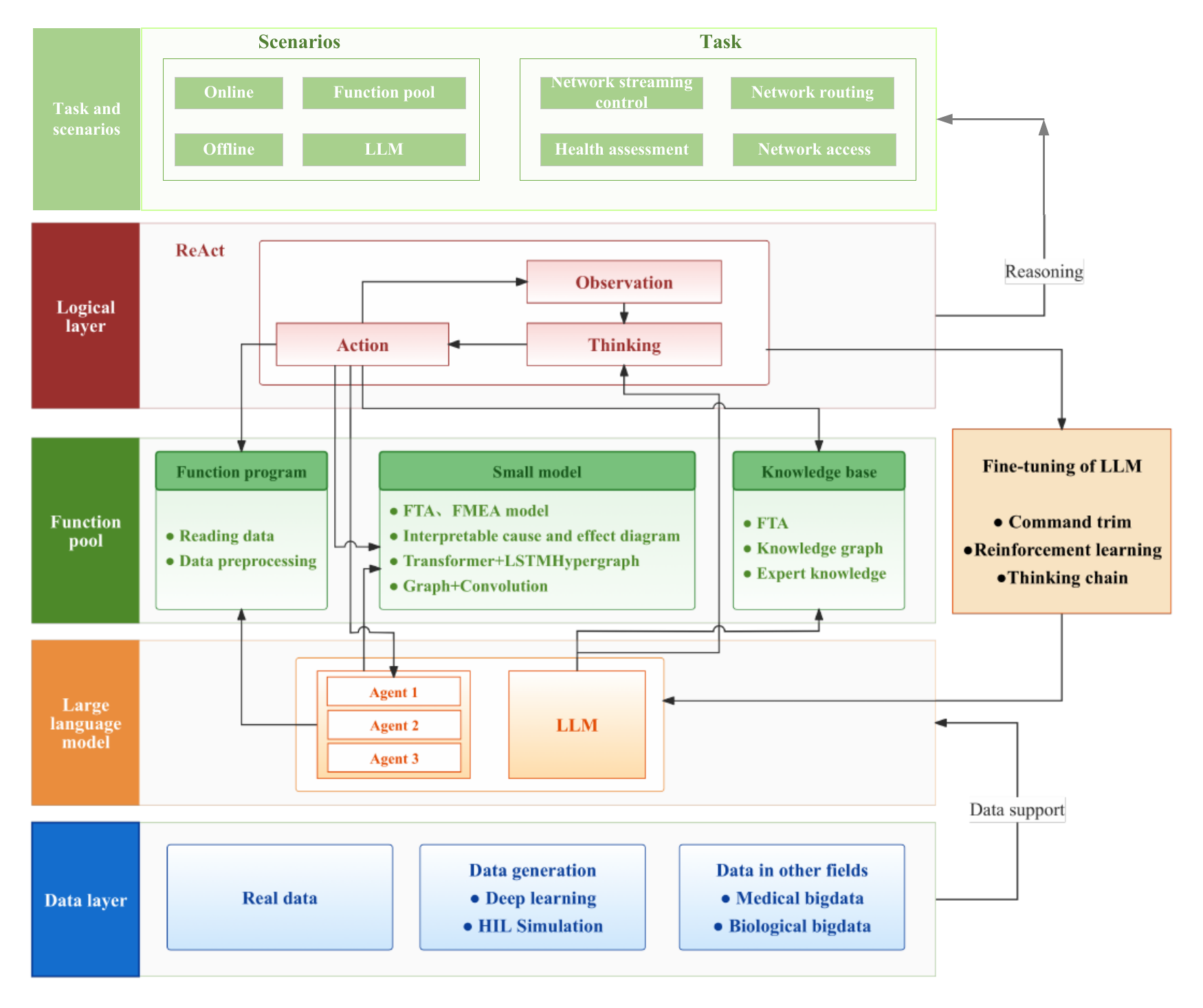}
	\caption{Overall design scheme for network health assessment and performance optimization architecture based on LLM.}
	\label{fig_c1}
\end{figure*}
Network operation and optimization face multiple technical barriers, which involve multiple levels, making the work of operation and optimization complex and challenging.
Network operation and optimization require continuous learning and adaptation to new technical standards and protocols, such as the 5G transition to 6G applications requiring updates to corresponding protocols. In addition, there are many important challenges in network operation and optimization, such as fault diagnosis and network health assessment. In a complex network environment, how to effectively monitor network performance, detect and locate faults in a timely manner is a huge challenge.
For example, in order to validate the application of LLM in network operation and optimization, as shown in Figure 3, this paper proposes an intelligent network architecture based on LLM for network intelligent operation and optimization. It mainly consists of several components, namely the Data layer, Large Language Model, Function pool, Logical layer, and Task and scene modeling.

These components serve as data, platform, logic layer, function pool, and model support to provide personalization services for the entire network operation system, such as centralized global diagnosis, distributed inference, network routing, network access and fault diagnosis, interpretable performance evaluation and key factor traceability, and task oriented health assessment. The functions and interrelationships of these components are as follows:
\begin{itemize}
	\item{\textbf{Data layer: }}
	Data resources serve as the cornerstone for Large Language Models to efficiently network operations and optimization. These resources can be primarily divided into three parts: 1) real data, 2) generated data, and 3) data from other fields. The construction and maintenance of the data layer involve data storage, management, and manipulation, providing effective data support for model training and inference. During model training, real data is obtained through sensors and corresponding data structures, such as physical layer faults, network layer related operation data, and data link layer fault data. These data undergo standardized processing procedures before being inputted into the LLM for effective model training. However, it is difficult to obtain a large amount of real data for training in real-world scenes. Therefore, efficient data generation methods become an important research focus. These methods utilize specific algorithms and techniques to simulate and generate data. For instance, Generative Adversarial Networks (GANs) [12] can produce high-quality text and highly realistic image data. This algorithm can also generate operation data that approximates reality in network operations, partially addressing the issue of insufficient training data. Additionally, as an important supplement for data acquisition, Hardware in the Loop (HIL) simulation is another effective data generation method. This approach combines key components and algorithms without fully deploying the entire system, simulating network operations to obtain data close to real ones. This data generation method offers real-time performance, scalability, and flexibility, making it adaptable to various simulation requirements and scenarios. It plays a crucial role in generating efficient 6G network operation data.
	\item{\textbf{Large Language Model: }}
	This module is primarily used to provide network fault detection algorithms and performance analysis tools in support of network intelligent operations. Based on differences in scale, number of parameters, computing power, adaptation scenarios, and functional flexibility, the model base can be further subdivided into small model bases and large model bases. 
	However, while the small model base can be flexibly and conveniently deployed on cloud-edge computing platforms, its network operation capabilities are limited by factors such as model accuracy, storage, data volume, and transmission rates. This makes it challenging to meet the service requirements of 6G networks, which demand high-speed, low-latency, high-traffic density, and comprehensive intelligence. Therefore, to achieve more intelligent and personalized network services, it is necessary to deploy a Large Language Model base to support efficient network operation and intelligent network operation services.
	Given the superior capabilities of LLM in processing multi-modal data such as text, images, and videos, they can easily learn and understand vast amounts of network operation data. For example, in dealing with network faults arising from network operation tasks, it is necessary to identify and deduce the types, causes, and impacts of the faults, and then recognize various fault modes. By analyzing real-time data from the operation of 6G networks, using LLM as an agent, it can also quickly identify potential or ongoing faults and provide timely warnings.
	When a fault occurs in the 6G network, the LLM can analyze relevant network logs, configuration information, performance data, and combine its learned knowledge to infer the possible cause of the fault. This helps to rapidly locate the problem and reduce the time required for fault investigation. Furthermore, based on a vast library of fault handling cases and expert experience, the LLM can recommend tailored solutions for specific faults. This not only provides timely assistance to network administrators but also optimizes solutions based on historical data, enhancing the efficiency of fault handling.
	Additionally, by analyzing historical network operational data and trends, LLM can predict potential future faults. This allows network administrators to take proactive measures to prevent faults and ensure the stable operation of 6G networks. By leveraging LLM to build a more robust network fault diagnosis system and continuously learning and updating its knowledge base through real-time data collection and analysis, LLM can improve the accuracy and efficiency of fault diagnosis ensuring the stability of network operations.
   \item{\textbf{Function pool: }}
   The functional pool mainly consists of three parts: 1) functional programs, 2) small models, and 3) knowledge bases. Among them, functional programs are primarily responsible for data reading and preprocessing tasks. For example, network fault data fed back by LLM needs to be output in a standardized format to the corresponding interface for invocation, making the architectural design of the entire network operation system more modular and intelligent.
   Secondly, the small model base mainly includes Fault Tree Analysis (FTA) and Failure Mode and Effects Analysis (FMEA) [13] for handling network operation. The FTA method can establish a set of top-down network fault diagnosis methods. Through fault tree analysis, it can understand the causes of various faults or failures in the network operation system and find the best way to reduce these risks. However, unlike the FTA method, the FMEA method starts with the reasons for the occurrence of events and identifies the potential consequences that may result from these events. For example, physical layer faults in a network may lead to issues such as data transmission interruptions, connection device failures, and mismatched transmission rates. 
   In network performance optimization, the hybrid model of Transformer+Long Short-Term Memory (LSTM) [14] is highly suitable for processing and predicting important events with extremely long time series intervals and delays. This holds significant practical implications for safeguarding the efficient communication services of 6G networks over extended periods of time. Given the self-attention mechanism of Transformer, it is capable of capturing global dependencies when processing sequential data and exhibits excellent parallel computing capabilities. Meanwhile, LSTM, as a variant of Recurrent Neural Networks (RNNs), is particularly adept at handling sequential data with long-term dependencies. By combining Transformer and LSTM, a more robust network operation and optimization model can be constructed to support the efficient provision of services by 6G networks.
   In addition, another important function of the functional pool is to establish a comprehensive knowledge base, including the construction of fault tree diagnosis analysis, a robust knowledge graph structure, and rich expert prior knowledge. For example, the knowledge graph provides a powerful semantic network by building comprehensive associations between network system failure modes, failure causes, failure handling measures, and other information. This allows the system to intelligently understand users' colloquial questions and return accurate answers. Fault information retrieval based on the knowledge graph can greatly improve the accuracy and efficiency of retrieval, helping operation personnel quickly locate and resolve network failures.
   \begin{figure*}[!t]
   	\centering
   	\includegraphics[width=7.5in]{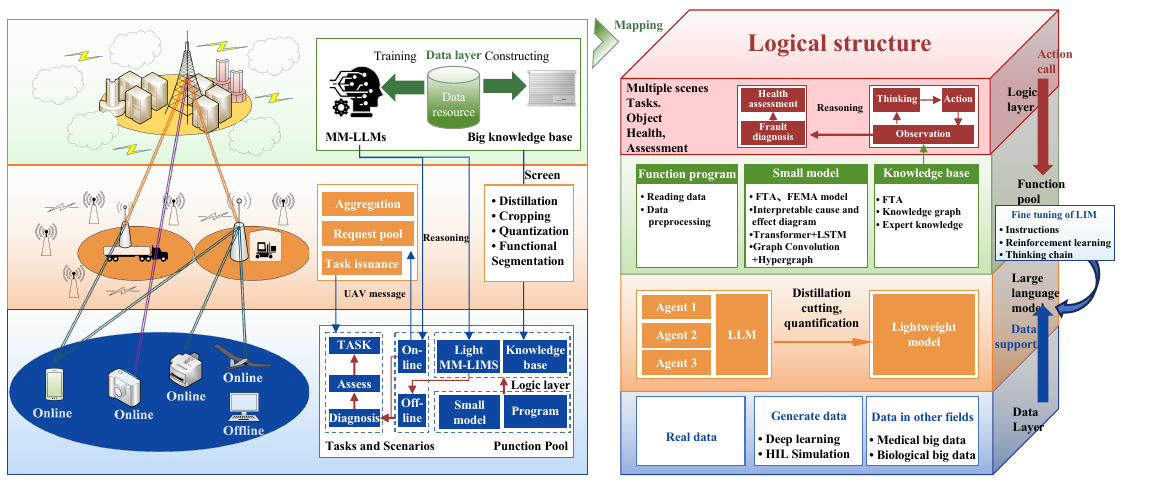}
   	\caption{Overall architecture diagram of an intelligent Cloud-Edge-Device collaborative Network Health Assessment System based on LLM.}
   	\label{fig_c2}
   \end{figure*}
   On the other hand, expert knowledge also plays an important role in network operation and optimization. For example, expert system technology is applied to network fault diagnosis, providing accurate fault diagnosis and maintenance solutions. This system can not only complete fault diagnosis within a specified time, but the expression of fault phenomena can usually guide the system builder to construct reasonable thinking more accurately, thereby improving the accuracy of fault diagnosis. In addition, the expert system can intelligently adjust parameters during system identification, achieving the goal of automated debugging.
   \item{\textbf{Logic layer: }}
   To design an efficient and modular network operation system, the logic layer plays a crucial role in the process of network operation, serving as an abstraction of the underlying structure. The logic layer structure comprises three core modules: observation module, thinking module, and action module. Among them, the thinking module is primarily responsible for integrating decision-making information from the underlying layers, such as network fault diagnosis information provided by large models and their corresponding solutions. This module plays a vital role in fault localization within the network health assessment system. Secondly, the action module translates the decisions made by the thinking module into actual action information. For instance, it provides data processing functions to the functional pool, enhances the expert prior knowledge base, and supplies action inputs to the LLM agent as feedback. Additionally, it can recommend appropriate solutions or emergency measures based on the type and severity of the fault, assisting users in rapidly restoring normal network operations. Consequently, the role of the logic layer in the network health assessment system is to analyze, diagnose, and process network fault information fed back from the underlying layers, while providing decision support and fault records. This helps users better manage and maintain the network, ensuring its stability and security. Furthermore, the logic layer provides reasoning services to upper-level application scenarios, thereby safeguarding the integrity of the entire health assessment system's functional structure.
   \item{\textbf{Tasks and scenarios: }}
   Tasks and scenarios are two important core modules of the application layer. Scenarios are mainly divided into online and offline states, which are interconnected with LLMs. The task module, on the other hand, is primarily responsible for operation and optimization of the entire network. It accomplishes network health assessment tasks, intelligent monitoring and network performance optimization by invoking relevant decisions from the logic layer. 
   Meanwhile, network operation scenarios are diverse, potentially involving hardware, software, configuration, security attacks, and various other situations. The poor operation of these scenarios can lead to issues such as network outages, increased latency, and elevated packet loss rates, ultimately affecting the normal provision of network services. 
   Furthermore, in network fault assessment, modularization and intelligentization can better handle the relationship between scenarios and tasks. Modularization breaks down the complex process of network operation assessment into multiple independent modules, with each module responsible for handling specific tasks, thereby improving the efficiency and accuracy of the assessment. Intelligentization, on the other hand, leverages LLMs to automatically identify and diagnose network faults, providing intelligent decision support for task issuance and execution in network operation and optimization scenarios.
\end{itemize}

Integrating the above functions,
as depicted in Figure 4, we designed a Cloud-Edge-Device intelligent network efficient operation system based on LLM, which is a novel intelligent network system that integrates LLM technology, cloud computing, edge computing, and terminal devices. Leveraging the strengths of LLMs in text understanding and generation, this system enables precise evaluation and effective management of network operation status.
Firstly, task assignments are issued at the edge of the network, where LLMs undergo pruning, distillation, and other optimization techniques. Various terminal devices, such as smartphones, cameras, and drones, collect real-time data related to network operation, including network latency, packet loss rate, and bandwidth utilization. These raw data undergo preliminary processing at the edge devices according to task requirements, including data cleaning, format conversion, and initial pre-training. This lays the foundation for subsequent analysis and evaluation, such as pre-training lightweight LLMs and establishing local knowledge bases.
Secondly, the cloud-based LLM also participates in the data processing, extracting relevant information related to network operation (eg. health status) by understanding the linguistic patterns and structural features within the textual data. It establishes a vast knowledge repository of network operation data, enabling the training of LLMs specifically tailored for network health conditions. For instance, it can diagnose the location of faults based on network link and protocol failure information and archive these details.
The cloud-based LLM, trained on vast amounts of textual data, possesses the capability to understand, generate, and process natural language. It receives data from edge devices, combines it with historical and real-time data, and conducts a thorough analysis of network health status. By comprehending the inherent patterns and potential semantic information within the network health data, the LLM can accurately identify potential issues, predict changes in network performance trends, and provide corresponding evaluation results.
Furthermore, the analysis and evaluation results of the LLM are presented to users through a visual interface. Users can view real-time information about the network's health status, historical trends, and potential risks. When the system detects abnormalities in the network health status, the LLM generates corresponding warning messages and provides decision support. These warnings and suggestions assist users in quickly locating issues and developing solutions, thereby enhancing the efficiency and accuracy of network management.

\section{Our demo}
\begin{figure*}[!t]
	\centering
	\includegraphics[width=7.5in]{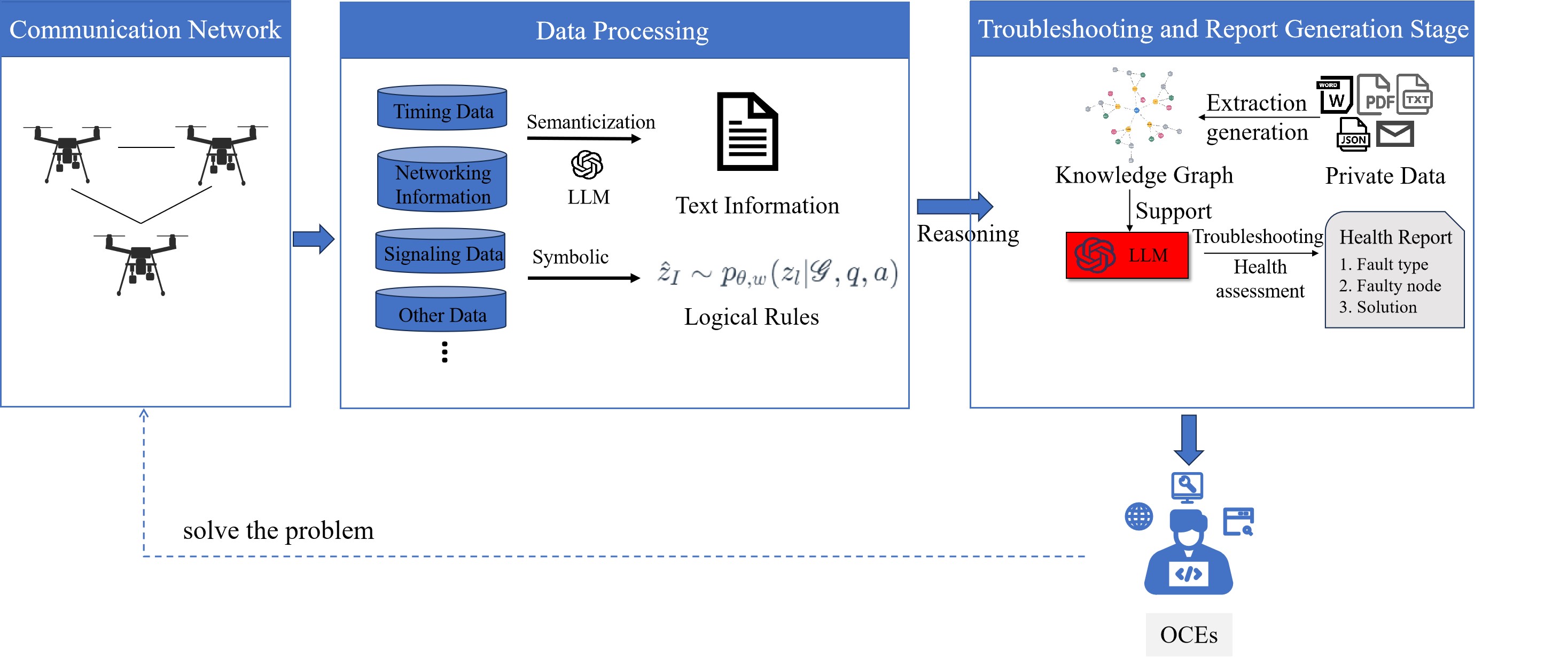}
	\caption{Schematic diagram of network health management system architecture based on textual proposition learning and LLM.}
	\label{fig_c3}
\end{figure*}
This section mainly demonstrates some of our current progress in the application of LLM in network health assessment. Nodes in highly dynamic heterogeneous networks are characterized by strong mobility and heterogeneity, resulting in frequent and complex network failures. To ensure the reliability and availability of network communications, as shown in Figure 5, this paper proposes a network health management scheme based on intelligent network information textualization, proposition learning, and LLM. Firstly, we extract a large number of labeled network failure datasets from the digital twin network, which are highly similar to real-world network conditions, to address the issue of insufficient fault samples. Then, by introducing a tree-structured semantic generation model, structured data is first converted into the required text sentences, mapping the information status of key network nodes to words representing the current indicator status. After collecting abnormal node information for a specific node, a fault report is generated through an enhanced retrieval and generation architecture. Private domain data is used to improve the input of model, generating descriptions with abnormal information and diagnostic methods. Finally, the output of the Large Language Model is used to update the tree structure, improving the model's accuracy and inference efficiency, and ultimately generating high-quality network health reports.

It is noteworthy that in the architecture diagram depicted in Figure 5, drones communicate information via a dedicated network, ensuring efficient real-time data transmission. Furthermore, time-series data incorporating network information is first processed through a large language model (LLM) for semantic understanding and conversion into textual information. Simultaneously, logical rules are employed to symbolize signal data, creating fused data from multiple modalities for subsequent LLM training. This data encompasses flight control information and various related parameters of drones, potentially containing fault samples as well. For instance, a circuit malfunction in a drone may trigger a series of chain reactions leading to data anomalies.

Moreover, a knowledge graph is utilized to extract and unify data from these different modalities, which is then fed into the LLM for training. Fine-tuning of the LLM is also conducted as necessary. Finally, the large model conducts health assessments based on the input data. It trains a fault discriminator through embedding models and can also leverage causal inference methods in conjunction with the LLM for troubleshooting. Subsequently, a health report is generated, outlining fault types, faulty nodes, and corresponding solutions, thereby achieving comprehensive network health assessment for the entire system.

In summary, the design and implementation of a drone health assessment system based on Large Language Models necessitate comprehensive consideration of multiple aspects, including data quality and bias, model training and validation, security and privacy protection, system architecture and integration, real-time performance and accuracy, user-friendliness, as well as compliance and ethical issues. By continually optimizing and refining these aspects, it can enhance the performance and reliability of the drone health assessment system, thereby providing robust support for the safe flight of drones.

\section{Several open issues}
With the continuous development and improvement of 6G networks, the application of Large Language Models within them will become more extensive and profound. However, we also face numerous challenges. On the one hand, as the scale and complexity of models increase, the costs of training and inference will also rise. On the other hand, it is an urgent issue to achieve efficient model operation and performance optimization while ensuring data privacy and security. Therefore, we need to continuously explore new technologies and methods to address these challenges. 
However, important issues will also be faced in the research of intelligent networks and performance optimization.
\subsection{Model generalization ability and robustness}
The network performance optimization system based on Large Language Models may have the problem of insufficient generalization ability. Although Large Language Models have strong learning capabilities, they may suffer from overfitting and insufficient generalization ability in some cases. This means that the model may not adapt well to new operation scenarios and tasks, limiting its effectiveness in practical applications. Moreover, the issue of robustness is also a key technology that needs to be urgently addressed. The network health assessment system based on Large Language Models needs to handle various exceptional situations, including unknown faults, system crashes, and so on. Large language models need to possess sufficient robustness to cope with these challenges.

\subsection{Data understanding and representation}
In the context of network operation and performance optimization leveraging LLMs, understanding the representation of multimodal data such as semantics, text, images, and videos is a complex yet crucial process that significantly contributes to providing decision support for network operation. However, there remain significant challenges in exploring and implementing universal methods for the effective fusion and comprehension of multimodal data through technical means such as data alignment and correlation, cross-modal feature representation, and semantic understanding and reasoning.

Different modalities of network data (e.g., text, images, videos) may exhibit temporal or spatial correspondences. To effectively fuse these data, algorithms with superior performance need to be designed for data alignment and correlation operations, ensuring that data from various modalities can complement and enhance each other.

Furthermore, to uniformly represent the diverse modalities of network data within the same feature space, innovative cross-modal feature representation techniques are required. These techniques convert data from different modalities such as text, images, and videos into shared feature vectors or representations, enabling further processing and analysis by LLMs.

Once a unified representation of multimodal data is achieved, a more powerful LLM framework with enhanced representation capabilities must be designed for semantic understanding and reasoning. This framework should be capable of comprehending semantic information in network log texts, identifying key elements and events in images and videos, and inferring associations and causal relationships among different modalities of network operation data.

\subsection{Multimodal alignment problem}
6G network pursues three-dimensional coverage. It needs to build a more complex and flexible network topology to meet the coverage requirements in different scenarios. However, the traditional Large Language Model is dominated by single text or image mode, which will face great difficulties in realizing multimodal fusion and alignment in 6G heterogeneous complex network.
\subsection{Selection and equilibrium of LLM and small model}
In the 6G network, how to efficiently select between Large Language Models and small models is a crucial task that requires a careful balance of factors such as model performance, resource requirements, application scenarios, as well as data privacy and security. Determining the appropriate model type based on specific needs and scenarios, and optimizing the deployment and inference processes of these models, has become an important research focus in order to fully leverage the advantages of 6G networks and enhance overall performance.
\subsection{Model self evolution problem}
The issue of model self-evolution in 6G networks combined with reinforcement learning and multi-agent systems represents a challenging and promising research direction. A crucial question lies in how to fully leverage the advantages of 6G networks, as well as the unique characteristics of reinforcement learning and multi-agent technology, to achieve efficient self-evolution of models. For instance, by integrating reinforcement learning with multi-agent systems, models can take advantage of the synergistic effects among multiple agents, enabling them to achieve more efficient self-evolution.
\subsection{Data collection and real-time performance of the system}
Large-scale data collection faces a series of issues, including the costs of data transmission and storage, as well as concerns regarding data accuracy, reliability, privacy, and security protection. At the same time, data security issues should not be ignored. For example, Mišić \textit{et al.} [15] discussed in detail a large number of existing security and privacy related restrictions on the network, which reflected the importance of network security in the context of Internet of things applications. To address these challenges, it is necessary to adopt advanced data compression and encryption technologies while leveraging cloud computing and edge computing techniques to achieve efficient data processing and utilization, this can effectively utilize the latest computing modes to solve the problems caused by data transmission. However, due to the complexity and uncertainty of network transmission, as well as the limitations of device processing capabilities, achieving system real-time performance is not an easy task. Therefore, one reliable approach is to incorporate data simulation and future prediction from digital twin networks to tackle this difficulty.
\subsection{Security and interpretability of LLM in networks}
The issues of security and interpretability of LLMs in 6G networks are significant challenges currently faced and expected to persist in the field of communication technologies. Due to their extensive parameters and complex structures, Large Language Models are susceptible to exploitation by attackers. Addressing data privacy and leakage concerns is paramount, necessitating enhanced security measures for these models, including the adoption of advanced encryption technologies to safeguard the secure transmission and storage of model parameters and data. On the other hand, the application of LLMs in 6G networks also poses certain challenges. Their complexity and black-box nature often render their decision-making processes and output results difficult for humans to comprehend and interpret. To enhance the interpretability of LLMs in 6G networks, technical approaches can be employed to increase model transparency. For instance, visualization techniques can be utilized to demonstrate the internal structure and decision-making processes of the models, enabling users to gain a more intuitive understanding of their operational mechanisms.

\section{Conclusion}
The LLM have broad application prospects in 6G network performance optimization and operation, which also face many challenges, such as model performance, data quality, privacy security, and interpretability issues. To overcome these challenges, 
this paper proposes a network intelligent operation and performance optimization framework based on LLM. By analyzing their role in network health assessment, fault detection, and other aspects, it is found that LLM have great potential for enhancing the intelligence level of 6G networks. Although the performance of LLM may be limited when processing complex network health assessment tasks, their advantage in processing natural language provides a new perspective for network state analysis. By learning and analyzing historical data, the model can predict the future trend of network traffic changes, providing decision-making basis for network resource allocation and service quality improvement. 
In the future outlook, with the continuous development of 6G technology, LLM will play a more important role in network performance optimization, fault prevention, and resource scheduling, providing strong support for building a more intelligent, efficient, and secure 6G network.

\section{Acknowledgments}
This work was supported by the Postdoctoral Fellowship Program of CPSF (Grant No.GZC20233160), Changsha Municipal Natural Science Foundation (Grant no.kq2208284), Hunan Provincial Natural Science Foundation (Grant no.2023jj40774), National Natural Science Foundation of China (Grant no.62302527), Hunan Provincial Natural Science Foundation Enterprise Joint Fund (Grant no.2024JJ9173). We sincerely appreciate the efforts made by the editor and reviewers to improve this paper.

\end{document}